# High-spin orbital interactions across van der Waals gaps controlling the interlayer ferromagnetism in van der Waals ferromagnets

Hyun-Joo Koo,[a] Reinhard K. Kremer,[b] and Myung-Hwan Whangbo*,[a,c]

[a]Department of Chemistry and Research Institute for Basic Sciences, Kyung Hee University, Seoul 02447, Republic of Korea
[b]Max Planck Institute for Solid State Research, Heisenbergstrasse 1, D-70569 Stuttgart, Germany
[c]Department of Chemistry, North Carolina State University, Raleigh, NC 27695-8204, USA.

ABSTRACT: We examined what interactions control the sign and strength of the interlayer coupling in van der Waals ferromagnets such as $Fe_{3-x}GeTe_2$, $Cr_2Ge_2Te_6$, $CrI_3$ and $VI_3$, to find that high-spin orbital interaction across the van der Waals gaps are a key to understanding their ferromagnetism. Interlayer ferromagnetic coupling in $Fe_3-xGeTe_2$ $Cr_2Ge_2Te_6$, and $CrI_3$ is governed by the high-spin two-orbital two-electron destabilization, but that in $VI_3$ by the high-spin four-orbital two-electron stabilization. These interactions explain a number of seemingly puzzling observations in van der Waals ferromagnets.

Due to their potential for spintronics and data storage applications, van der Waals (vdW) ferromagnets attracted broad attention recently. Mostly, they are layered tellurides (e.g., $Fe_{3-x}GeTe_2$ (FGT)[1-5] and $Cr_2Ge_2Te_6$[6]) and layered iodides (e.g., $CrI_3$[7-9] and $VI_3$[10,11]) composed of individually ferromagnetic (FM) monolayers. Practical applications of these materials will require use of thin slabs of several monolayers, so it is desirable that such slabs have a high Curie temperature $T_C$. In advancing toward this goal, it is crucial to know what controls the interlayer coupling to be FM or antiferromagnetic (AFM) and how one might enhance the strength of the interlayer FM coupling. In FGT (Figure 1), the interactions between adjacent layers are determined primarily by those between two adjacent sheets of Te atoms, and in $CrI_3$ and $VI_3$ by those between two adjacent sheets of I atoms. vdW interactions cannot be responsible for whether the interlayer interactions are FM or AFM, because they are weak, fall off very rapidly with distance, and are independent of spins.

As the spin exchanges between adjacent layers of FGT, one may consider Fe-Te…Te-Fe exchanges across the vdW gaps. These exchanges, being determined by the interlayer Te…Te overlap across the vdW gap, are typically AFM rather than FM.[12] Thus, they cannot be responsible for the interlayer FM coupling in FGT. Numerous experimental studies found that FGT samples exhibiting ferromagnetism have Fe-vacancies,[1,2,13-15] and DFT calculations showed[5] that, only when FGT samples have a large enough hole concentration, their interlayer interaction becomes FM. However, a simple conceptual picture explaining why this is the case has not emerged yet.

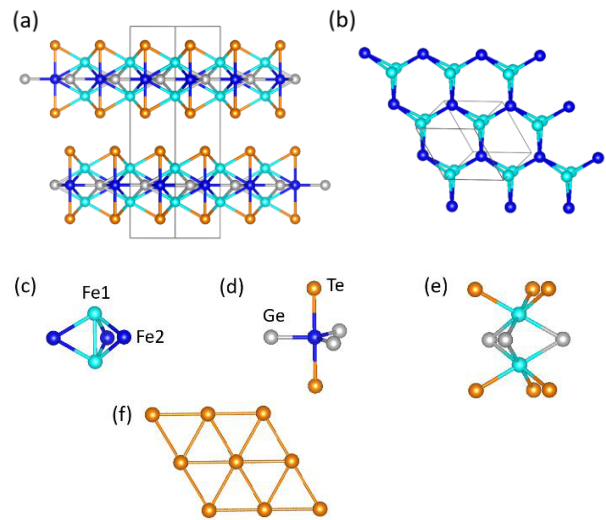

FIGURE 1. Crystal structure of $Fe_{3-x}GeTe_2$ and its building units. (a) A projection view of the crystal structure showing two layers per unit cell. (b) A perspective view of the $Fe1_2Fe2$ layer. (c) $Fe1_2Fe2_3$ trigonal bipyramid. (d) $Fe2Ge_3Te_2$ trigonal bipyramid. (e) Two $Fe1Ge_3Te_3$ octahedra sharing their $Ge_3$ faces. (f) A trigonal sheet of Te atoms. The Fe1, Fe2, Ge and Te atoms are defined in (c) and (d).

Fascinating but puzzling observations reported from electric double layer gating (EDLG) experiments[4] on a thin FGT slab remained unexplained, so far. In the EDLG experiments performed as a function of the gate voltage, $0 \le V_g \le 3V$, (for a schematic diagram, see Figure S1 in the Supporting Information, SI), the $T_C$ of the apparently substantially hole-doped FGT sample is ~100 K at $V_{g_r} = 0$ (cf. $T_C \approx$

220 K for bulk FGT). With increasing $V_g$ to ~1.5 V, $T_C$ sharply jumps to ~300 K. This was ascribed[5] to electron-doping induced by the gate voltage. Since holes are already present in the FGT sample, electron-doping would reduce the hole concentration in the sample so that one might have expected a decrease, rather than a sharp increase in $T_C$. (Alternatively, the sharp $T_C$ increase was suggested to originate from a $Fe_{2-x}Ge$ impurity produced by the decomposition of FGT.[16]) Because, at present, it is unknown what interlayer interactions govern the strength of the interlayer FM coupling and hence the $T_C$ of vdW ferromagnets, it is difficult to explain experimental and computational results accumulated on vdW ferromagnets and hence predict new kinds of vdW ferromagnets. In this Communication, we identify those interactions necessary for understanding vdW ferromagnets in general by studying FGT as a representative example.

For our discussion in the following, it is crucial to recall important structural features of FGT. The layered structure of FGT with no Fe-deficiency, namely, FGT (x = 0), is depicted in Figure 1a-b. FGT has two nonequivalent Fe atoms, Fe1 and Fe2 (Figure 1c). Each Fe2 forms a $Fe2Ge_3Te_2$ trigonal bipyramid (Figure 1d), and each Fe1 a $Fe1Ge_3Te_3$ octahedron (Figure 1e). Each $FeGe_3Te_3$ octahedron is axially-compressed with Fe1-Te = 2.661 Å, Fe1-Ge = 2.655 Å, and ∠Te-Fe1-Te = 97.76°. Each $FeGe_3Te_2$ trigonal bipyramid has shorter Fe-Te and Fe-Ge bonds, namely, Fe2-Te = 2.613 Å and Fe2-Ge = 2.314 Å. Thus, the space around each Fe2 is essentially smaller than that around Fe1, showing why the Fe-vacancies occur at the Fe2 sites.[1]

A clue to the interlayer interactions needed for describing the interlayer ferromagnetism is found from the EDLG experiments[4] with $LiClO_4$ as electrolyte. With the positive potential of $V_g$ acting on the gate electrode, the $Li^+$ ions intercalate in between the adjacent Te sheets (Figure S1 in the Supporting Information, SI), i.e., at the $Te_6$ octahedral sites of the vdW gaps, and interact with the filled p-orbitals of the surrounding $Te^{2-}$ ions to lower the energies of the Te p-states. Such cation-anion interactions are stabilizing in nature,[17] and will effectively weaken the direct interlayer Te...Te overlap across the vdW gaps because the Te p-orbitals are tied up with the cation-anion interactions. In short, the interlayer interaction we search for has something to do with interlayer orbital interactions. This conclusion is further supported by another piece of experimental evidence; $T_C$ of bulk FGT monotonically decreases with pressure, and its ferromagnetic ordering is suppressed above 13.9 GPa.[18] An obvious structural change under pressure is to reduce the vdW gaps, which enhances the interlayer Te...Te overlap. The latter disfavors interlayer FM coupling, or, alternatively, favors interlayer AFM coupling.

The interlayer interaction that can explain the above observations is the high-spin two-orbital two-electron (2O2E) interactions between two spin sites (Figure 2).[19,20] Consider a magnetic dimer of spin sites 1 and 2, described by magnetic orbitals $\phi_1$ and $\phi_2$, respectively. In the one-electron picture neglecting electron spins, the interaction between the magnetic orbitals create the bonding and antibonding states, $\psi_1$ and $\psi_2$, respectively, when the overlap $S = \langle\phi_1|\phi_2\rangle$ is nonzero (Figure 2a). The energy-lowering $\Delta_1$ of $\psi_1$ is smaller in magnitude than the energy-raising $\Delta_2$ of $\psi_2$ (i.e., $\Delta_2 > \Delta_1$).[19] Let us now take electron spin into consideration. For a FM arrangement between two spin sites possessing one magnetic orbital each, the result of the interaction is a high-spin 2O2E destabilizing interaction (Figure 2b, left), because $\Delta_2 > \Delta_1$. The extent of this destabilization, $\Delta E = \Delta_2 - \Delta_1$, is proportional to $S^2$ (i.e., $\Delta E \propto S^2$).[19] Thus, the larger the overlap S, the greater the destabilization $\Delta E$. To avoid such destabilization, the two spin sites may adopt an AFM arrangement because S = 0 in this case, hence incurring no 2O2E destabilization (Figure 2b, right).

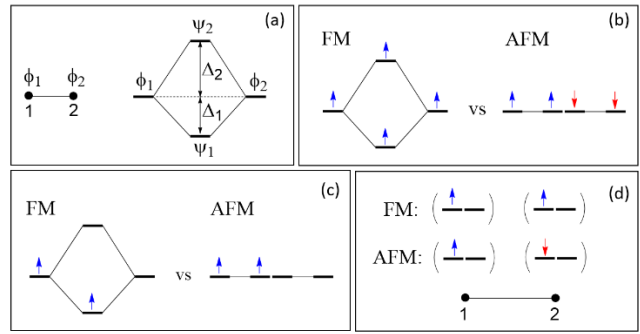

FIGURE 2. (a) Two sites of a dimer each possessing one orbital and one electron leading to the bonding and antibonding states in the one-electron picture, where $\Delta_1 < \Delta_2$ (electron spin neglected). (b) 2O2E interactions associated with the FM and AFM arrangements. (c) 2O1E interactions associated with the FM and AFM arrangements. (d) The FM and AFM arrangements of spin sites each possessing two magnetic orbitals and one electron leading to 4O2E interactions. Up-spin and down-spin electrons are represented by blue and red arrows, respectively.

Note that a FM arrangement can be more stable than an AFM arrangement if there is only one electron per two spin sites. For the FM arrangement provides a two-orbital one-electron (2O1E) stabilization (Figure 2c, left) whereas the AFM arrangement does not (Figure 2c, right). Such a situation arises, when each spin site has two magnetic orbitals but only one electron (Figure 2d). The high-spin four-orbital two-electron (4O2E) interaction resulting from this case leads to one 2O2E interaction and two 2O1E interactions. Since $\Delta_1 >> \Delta_2 - \Delta_1$ in general, the two 2O1E stabilizing interactions dominate over one 2O2E destabilizing interaction, hence favoring the FM arrangement over the AFM arrangement. This situation becomes relevant for the vdW ferromagnet $VI_3$ (see below).

FGT is a magnetic metal, rather than a magnetic insulator. Thus, we extend the concept of the high-spin 2O2E interaction for extended solids. In the spin-polarized description of a magnetic solid, each spin site has up- and down-spins in unequal amounts (Figure 3a). The majority spin (commonly, up-spin) is identical at all spin sites in a

FM arrangement but alternates between up-spin and down-spin from one site to another (Figure 3a) in an AFM arrangement. Suppose that each site of a magnetic solid has a majority spin state, as depicted in Figure 3b (middle). In the FM state, such majority spins interact to form a majority spin band (Figure 3b, left). In the AFM state, the majority spin of one site interacts with the minority spins of its neighboring sites, to form a majority spin band (Figure 3b, right). (See Figure S2 in the SI for further discussion, where we assumed for simplicity that the majority and minority bands of a given FM or AFM state have an identical band width although the minority band is somewhat wider than the majority band.[21]) Thus, the majority band of the FM state has a greater band width than does that of the AFM state. This simplified description remains valid for FGT, because each FM layer of FGT can be regarded as an effective single-spin entity. Thus, the majority spin band of FGT has a wider band, more antibonding in the upper region of the band and more bonding in the lower region of the band, in the FM state than in the AFM state. This provides a natural explanation for why FGT must have Fe-vacancies beyond a certain amount for it to be FM. The electronic states around $E_F$ are least stable, and the occupation of these high-energy states is avoided by hole-doping. The energy lowering by hole-doping is greater for the FM state than for the AFM state, because the FM state has more antibonding in the upper region of the band and more bonding in the lower region of the band.

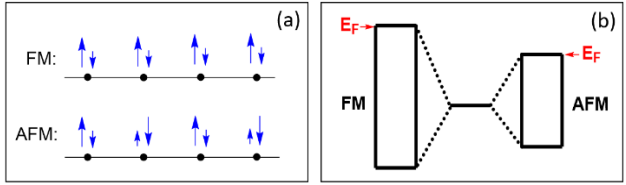

FIGURE 3. (a) Polarized spin description of a magnetic chain with FM or AFM spin arrangement. (b) Majority-spin bands expected when the spin sites of a uniform chain have FM and AFM arrangements.

The predictions of the conceptual picture are readily verified by analyzing how the relative energies of the FM and AFM states of stoichiometric FGT (i.e., x = 0) depends on hole- and electron-doping. We simulate this dependency by performing spin-polarized DFT calculations (details in the SI)[22-25] for FGT (x = 0) as a function of the total number of valence electrons $N_{tot}$ per formula unit (FU). For FGT (x = 0), $N_{tot}$ = 50 per FU. Since each Fe has eight valence electrons, the carrier density δ (per FU) introduced by calculations with $N_{tot}$ electrons per FU is given by δ = (50 − $N_{tot}$)/8. In this definition, hole-doping is represented by δ > 0, and electron-doping by δ < 0. The total and projected density of states (DOS) plots for FGT (δ = 0) are presented in Figure S3. The relative energies of the AFM and FM states, ΔE(δ) = $E_{AFM}$(δ) − $E_{FM}$(δ), calculated as a function of δ (Figure S4a and Table S1) show that the interlayer FM coupling is more stable than the interlayer AFM coupling only when the hole-doping is substantial (i.e., if δ > ~0.07). The ΔE(δ) vs. δ plot predicts that electron-doping makes the interlayer AFM coupling more stable than the interlayer FM coupling. These results are consistent with the qualitative picture discussed above. The total spin moment $\mu_{tot}$(δ) per FU calculated as a function of δ (Table S1) decreases as the extent of either hole- or electron-doping increases (Figure S4b and Table S1). To gain further insight into how hole/electron doping affects the interlayer interaction, we calculated the charge density plots of FGT for δ = 0, 0.1875 and -0.1875 (Figure S5 and S6). They show that hole (electron) doping decreases (increases) the electron density around the Te atoms and hence reduces (enhances) the interlayer overlap between Te 5p orbitals across the vdW gap, consequently promoting the interlayer FM (AFM) coupling. This finding is in full support of our conclusion based solely on energy considerations.

Another vdW ferromagnet $VI_3$ consists of honeycomb layers made up of edge-sharing $VI_6$ octahedra (Figure S7). Contrary to the case of FGT, the $T_C$ of $VI_3$ shows a monotonic increase with pressure, reaching $T_C$ ≈ 98 K at 6.3 GPa.[26] The $V^{3+}$ ($d^2$, S = 1) ion of $VI_3$ has the electron configuration $(1a)^1(1e)^1$, in which the doubly-degenerate level 1e has only one electron.[11] Thus, if only the 1e state of each $V^{3+}$ ion is considered, for simplicity, the interaction between two adjacent $V^{3+}$ ions corresponds to the high-spin 4O2E interaction (Figure 2c, d), which favors the FM over the AFM arrangement. The tendency for the interlayer FM coupling is enhanced by an increase in the interlayer I…I overlap under pressure, explaining why the $T_C$ of $VI_3$ increases strongly under pressure. Such a pressure-dependence of $T_C$ is not found for $CrI_3$ because the $Cr^{3+}$ ($d^3$, S =3/2) ion with $(1a)^1(1e)^2$ electron configuration[11] does not lead to a high-spin 4O2E interaction. In another vdW ferromagnet $Cr_2Ge_2Te_6$ containing $Cr^{3+}$ ions, the $T_C$ is found to decrease under pressure.[27] Finally, we note that the $T_C$ of $NaFe_{2.78}GeTe_2$ is very similar that of $Fe_{2.78}GeTe_2$ despite that the Na-intercalation causes one full electron-doping per $Fe_{2.78}GeTe_2$.[16] This reflects that the intercalated $Na^+$ cations reduce the direct interlayer Te…Te overlap across the vdW gaps hence weakening the interlayer AFM coupling.

In summary, the interlayer high-spin 2O2E interaction is responsible for the interlayer ferromagnetism of FGT, $CrI_3$ and $Cr_2Ge_2Te_6$, and the high-spin 4O2E interaction for that of $VI_3$. High-spin orbital interactions are a key to understanding the interlayer ferromagnetism in vdW ferromagnets.

## ASSOCIATED CONTENTS

Supporting Information. Figure S1 for a schematic view of the EDLG experiment, Figure S2 for a spin-polarized description of the FM and AFM states, Figure S3 for DOS plots calculated for FGT (x = 0), Figure S4 and Table S1 for the relative energy of the FM and AFM states and the total magnetic moment per FU, Figure S5 and S6 for charge density plots, Figure S7 for a crystal structure of $VI_3$, and Details of calculations.

## AUTHOR INFORMATION

### Corresponding Author

* mike_whangbo@ncsu.edu.

### Author Contributions

The manuscript was written through contributions of all authors.

## ACKNOWLEDGMENT

(The work at KHU was supported by the Basic Science Research Program through the National Research Foundation of Korea (NRF) funded by the Ministry of Education (2020R1A6A1A03048004).

## ABBREVIATIONS

FGT, $Fe_{3-x}GeTe_2$; FU, formula unit; FM, ferromagnetic; AFM, antiferromagnetic; DOS, density of states; 2O2E, two-orbital two-electron; 2O1E, two-orbital one-electron; 4O2E, four-orbital two-electron

SYNOPSIS TOC

The origin of the interlayer ferromagnetism in van der Waals ferromagnets was probed by analyzing the structural, electronic and magnetic properties of $Fe_{3-x}GeTe_2$. High-spin orbital interactions across the van der Waals gaps are a key to understanding the interlayer ferromagnetism.

- **Ferromagnetism in van der Waals ferromagnets**
- **High-spin orbital interactions across van der Waals gaps**
- **2O2E, 2O1E & 4O2E interactions**

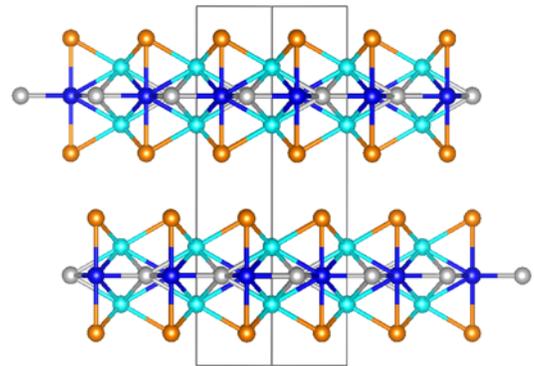

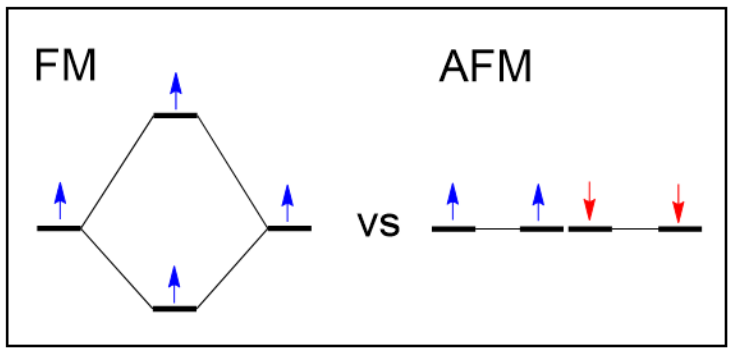

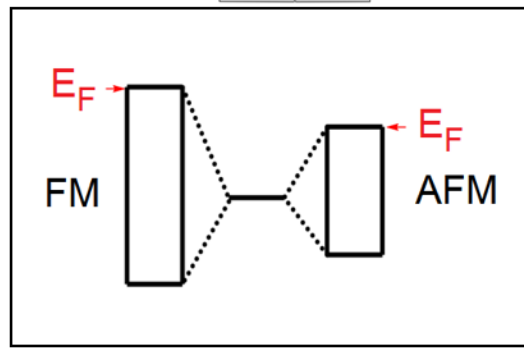

Supporting Information

for

**High-spin orbital interactions across van der Waals gaps controlling the interlayer ferromagnetism in van der Waals ferromagnets**

Hyun-Joo Koo,[a] Reinhard K. Kremer,[b] and Myung-Hwan Whangbo*,[a,c]

[a] Department of Chemistry and Research Institute for Basic Sciences, Kyung Hee University, Seoul 02447, Republic of Korea

[b] Max Planck Institute for Solid State Research, Heisenbergstrasse 1, D-70569 Stuttgart, Germany

[c] Department of Chemistry, North Carolina State University, Raleigh, NC 27695-8204, USA.

Email: mike_whangbo@ncsu.edu

## S1. Supplementary figures

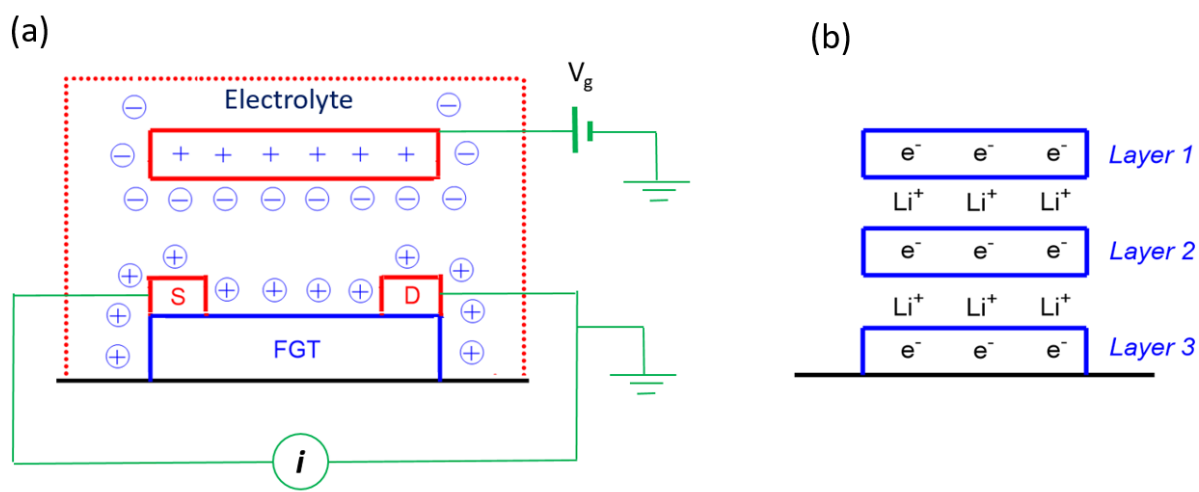

Figure S1. (a) Schematic view of the electric double layer gating (EDLG) experiment on a thin FGT slab using LiClO$_4$ as an electrolyte.[1] The positive gate voltage was applied on the electrode pushes the Li$^+$ ions toward the FGT slab, which induces Li$^+$-intercalation into the vdW gaps of the FGT slab and electron-doping into FGT layers. The resistance of the FGT slab is measured between the source (S) and the drain (D) by applying a voltage difference between the two. (b) Intercalation of Li$^+$ ions into the vdW gaps between adjacent FGT layers.

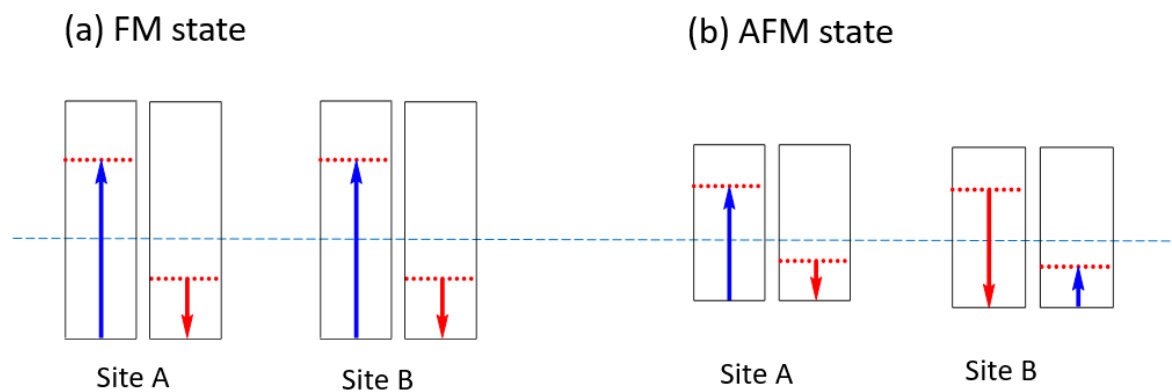

Figure S2. Spin-polarized description of the FM and AFM states, which are expected for a uniform chain of identical spin sites, (AB)$_\infty$. The FM (AFM) state is obtained when the adjacent spin sites are ferromagnetically (antiferromagnetically) ordered. The up-spin and down-spin bands are represented by blue and red arrows, respectively, the Fermi level indicated by the red dashed line. The majority spin is identical at all spin sites in the FM state, but alternates between up-spin and down-spin in the AFM state. At each spin site, the majority and minority spin band should be shifted to have an identical Fermi level. For the clarity of our discussion, this shift is suppressed. The blue dashed line represents the energy level of a single spin site. It assumed for simplicity that the majority and minority bands of a given FM or AFM state have an identical band width although the minority band is somewhat wider than the majority band.[2] This does not affect our conclusion, because what matters for the energy-lowering is the removal of electrons near the Fermi level of the majority band.

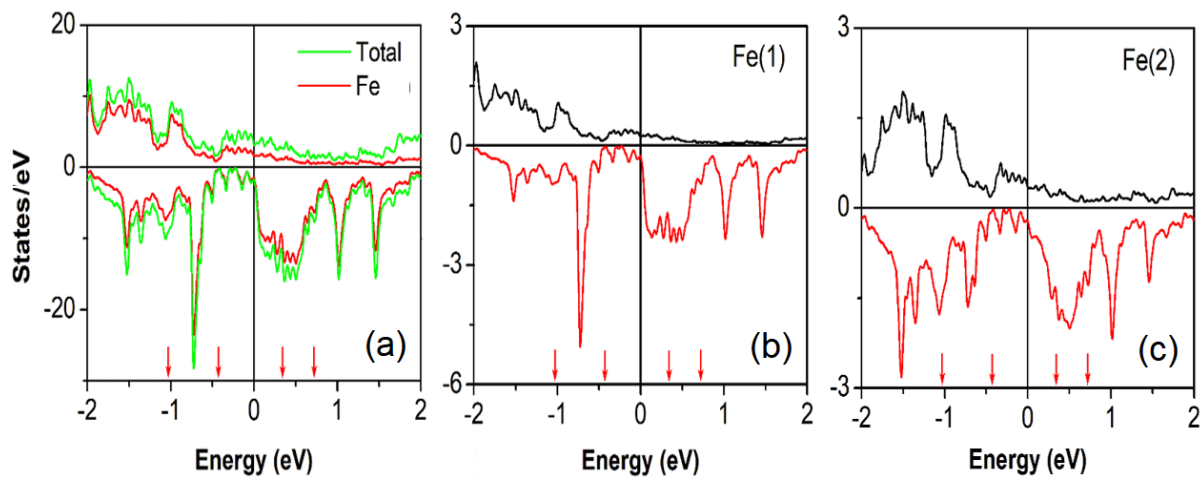

Figure S3. (a) The TDOS plot calculated for FGT (x = 0) and the PDOS plot for the 3d states of all Fe atoms. The PDOS plots calculated for the 3d states of the Fe(1) and Fe(2) atoms in (b) and (c), respectively. The small red arrows, from left to right on the energy axis, refer to the energies corresponding to δ = 0.25. 0.125, -0.125 and -0.25, respectively, and they occur at -1.036, -0.457, +0.355 and +0.695 eV, respectively.

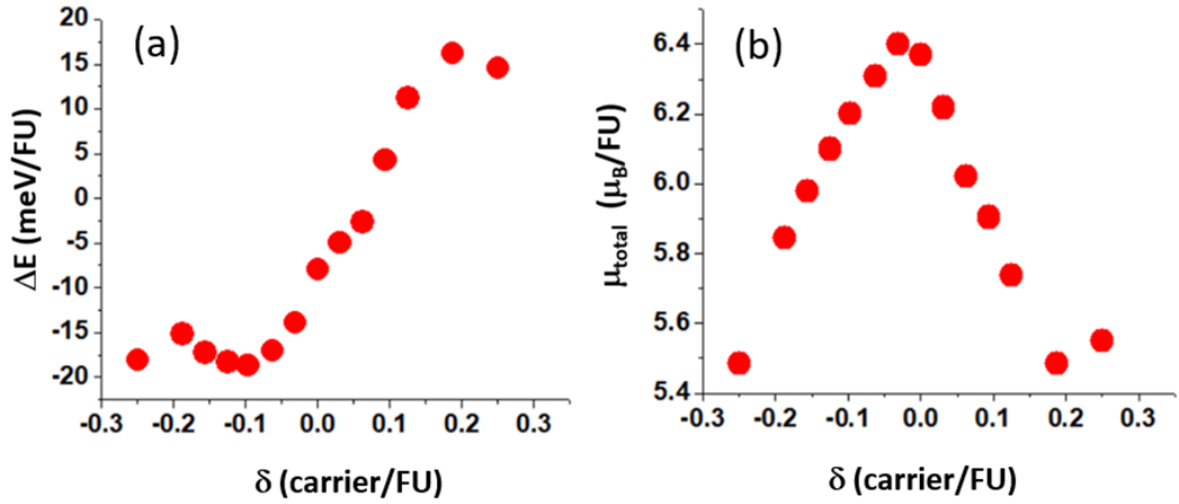

Figure S4. (a) The relative energy of the FM and AFM states, $\Delta E(\delta) = E_{AFM}(\delta) - E_{FM}(\delta)$, calculated for stoichiometric FGT(x = 0) as a function of the carrier density $\delta = (50 - N_{tot})/8$. (b) The total magnetic moment $\mu_{tot}(\delta)$ per FU calculated as a function of $\delta$.

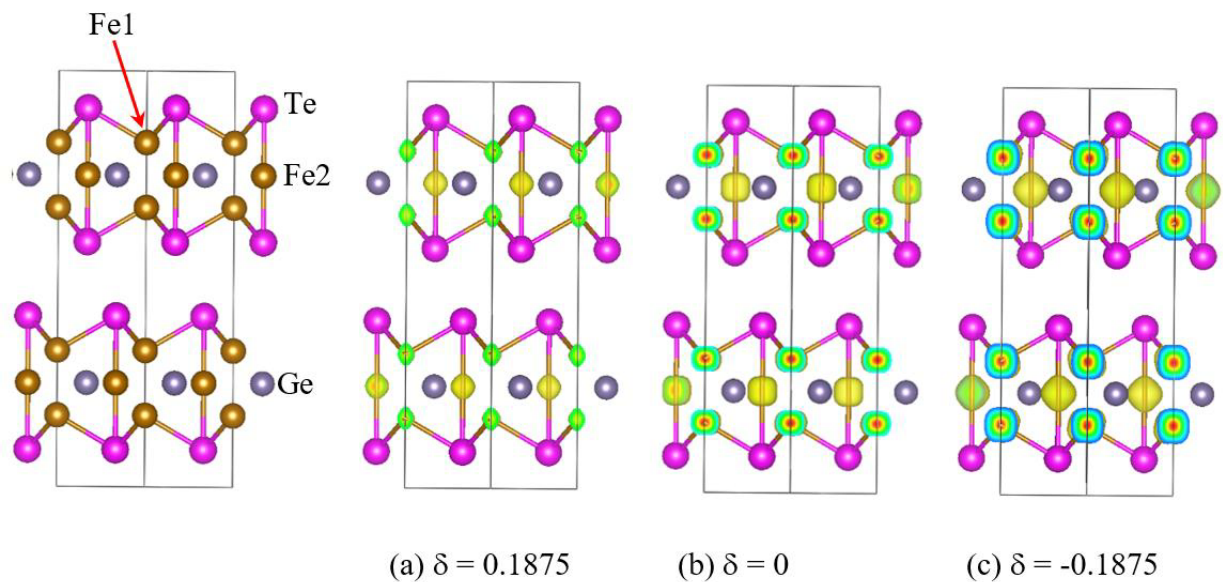

Figure S5. Surface contour plots calculated for the Te atoms of $Fe_{3-\delta}GeTe_2$ using the states lying within 0.05 eV from the Fermi level (surface contour value = 0.003): (a) $\delta = 0.1875$, (b) $\delta = 0$ and (c) $\delta = -0.1875$.

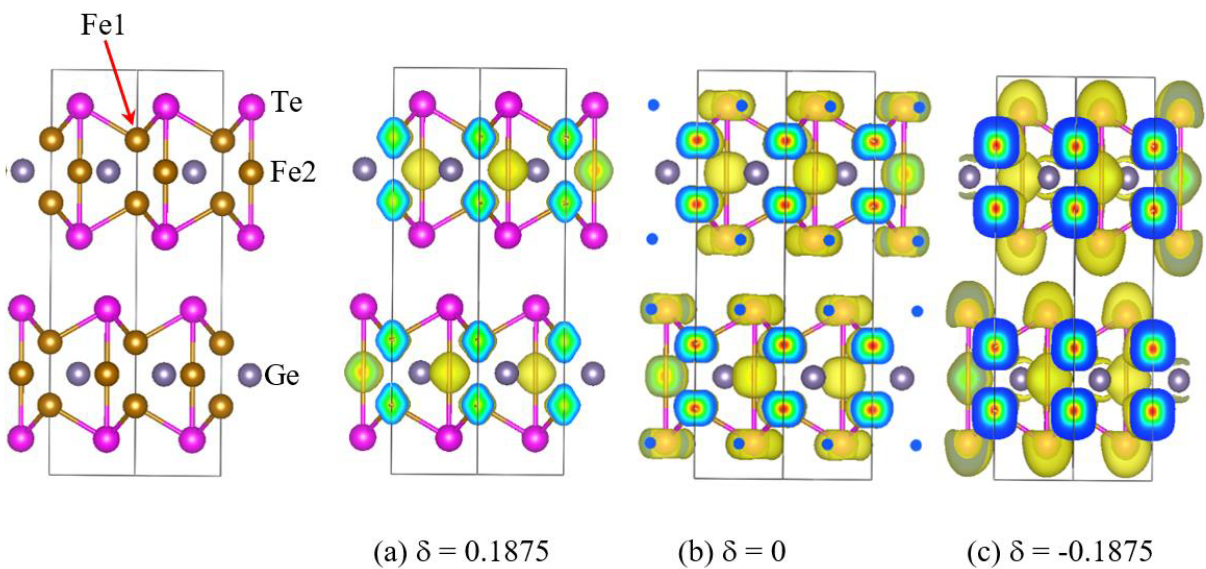

(a) δ = 0.1875     (b) δ = 0     (c) δ = -0.1875

Figure S6. Surface contour plots calculated for the Te atoms of $Fe_{3-\delta}GeTe_2$ using the energy states lying within 0.05 eV from the Fermi level and the surface contour value of 0.0005: (a) δ = 0.1875, (b) δ = 0 and (c) δ = -0.1875.

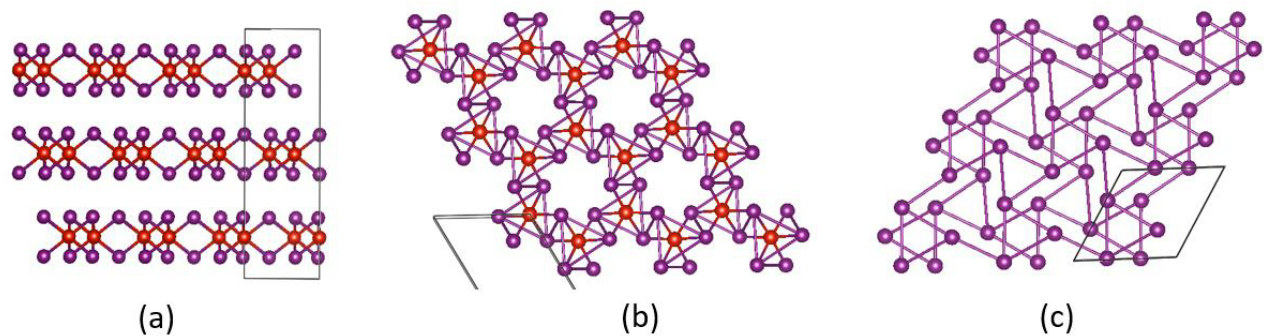

Figure S7. The crystal structure of $VI_3$ determined at 40 K.[3] (a) A projection view showing three layers per unit cell. (b) A projection view of a single $VI_3$ layer. (c) A projection view of two adjacent sheets of I atoms between two adjacent layers in $VI_3$. The shortest I…I distance is 3.729 Å between the two sheets of I atoms within a layer, and 4.104 Å between the two sheets of I atoms between adjacent layers.

## S2. Supplementary table

Table S1. The values of $\Delta E(\delta) = E_{AFM}(\delta) - E_{FM}(\delta)$ in meV per FU simulated by doing spin-polarized DFT calculations for stoichiometric $Fe_3GeTe_2$ with varying the total number of valence electrons $N_{tot}$ per FU, where the carrier density per FU is defined by $\delta = (50 - N_{tot})/8$. The total moment $\mu_{tot}$ ($\mu_B$) per FU as well as the moments on the Fe(1) and Fe(2) atoms are also summarized.

| $N_{tot}$ per FU | $\delta$ | $\Delta E$ (meV/FU) | $\mu_{tot}$ ($\mu_B$) | $\mu_s$ ($\mu_B$) per Fe(1) | $\mu_s$ ($\mu_B$) per Fe(2) |
|---|---|---|---|---|---|
| 52 | -0.25 | -18.09 | 5.49 | 2.04 | 1.44 |
| 51.5 | -0.1875 | -15.19 | 5.84 | 2.15 | 1.56 |
| 51.25 | -0.1563 | -17.26 | 5.98 | 2.21 | 1.58 |
| 51 | -0.125 | -18.28 | 6.10 | 2.26 | 1.59 |
| 50.75 | -0.09375 | -18.68 | 6.20 | 2.31 | 1.59 |
| 50.5 | -0.0625 | -17.04 | 6.31 | 2.37 | 1.59 |
| 50.25 | -0.03125 | -13.9 | 6.40 | 2.42 | 1.59 |
| 50 | 0 | -7.96 | 6.37 | 2.43 | 1.58 |
| 49.75 | 0.03125 | -5.00 | 6.22 | 2.42 | 1.53 |
| 49.5 | 0.0625 | -2.65 | 6.02 | 2.39 | 1.46 |
| 49.25 | 0.09375 | 4.30 | 5.90 | 2.39 | 1.40 |
| 49 | 0.125 | 11.18 | 5.74 | 2.38 | 1.30 |
| 48.5 | 0.1875 | 16.16 | 5.48 | 2.37 | 1.17 |
| 48 | 0.25 | 14.56 | 5.88 | 2.44 | 1.18 |

Table S2. Pseudopotential parameters employed for the Fe, Ge and Te atoms in DFT calculations.

| | $Z_{val}$ (e$^-$) | RWIGS (Å) | Valence |
|---|---|---|---|
| Fe | 8 | 1.302 | $d^7s^1$ |
| Ge | 14 | 1.217 | $d^{10}s^2p^2$ |
| Te | 6 | 1.535 | $s^2p^4$ |

## S3. Details of DFT calculations

Spin-polarized DFT calculations were carried out for the FM and AFM states of $Fe_3GeTe_2$ using the observed crystal structure for the two states without optimization. The interlayer coupling is FM in the FM state, and AFM in the AFM state. Our calculations emloyed the frozen core projector augmented plane wave (PAW)[4,5] encoded in the Vienna ab Initio Simulation Packages (VASP)[6] and the PAW_PBE potential (potpaw_PBE.5.4)[7] for the exchange-correlation functional. All our DFT calculations used the plane wave cutoff energy of 450 eV, a set of (21×21×9) k-points, and the threshold of $10^{-6}$ eV for self-consistent-field energy convergence. Our pseudopotential calculations employed the following parameters for the Fe, Ge and Te atoms are summarized in Table S2.